\documentclass[twocolumn,aps,prb,unsortedaddress,superscriptaddress]{revtex4}
\input epsf

\usepackage{graphicx}
\usepackage{dcolumn}
\usepackage{bm}
\usepackage{color}%
\usepackage{ulem}%

\begin{document}

\title{Extended H\"uckel theory for bandstructure, chemistry, and transport. \\ II. Silicon}

\author{D. Kienle}
\affiliation{Purdue University, Department of Electrical and Computer Engineering, West Lafayette, IN 47907, USA}
\author{J.I. Cerda}
\affiliation{Instituto de Ciencia de Materiales de Madrid, CSIC, Cantoblanco 28049, Madrid, Spain}
\author{K.H. Bevan}
\affiliation{Purdue University, Department of Electrical and Computer Engineering, West Lafayette, IN 47907, USA} 
\author{G-C. Liang} 
\affiliation{Purdue University, Department of Electrical and Computer Engineering, West Lafayette, IN 47907, USA}
\author{L. Siddiqui} 
\affiliation{Purdue University, Department of Electrical and Computer Engineering, West Lafayette, IN 47907, USA}
\author{A.W. Ghosh}
\affiliation{University of Virginia, Department of Electrical and Computer Engineering, Charlottesville, VA 22903, USA}
\date{\today}

\widetext
\begin{abstract}
In this second paper, we develop transferable semi-empirical parameters for
the technologically important material, silicon, using Extended H\"uckel 
Theory (EHT) to calculate its electronic structure. The EHT-parameters are
optimized to experimental target values of the band dispersion of bulk-silicon.
We obtain a very good quantitative match to the bandstructure characteristics 
such as bandedges and effective masses, which are competitive with the values 
obtained within an $sp^3 d^5 s^*$ orthogonal-tight binding model for 
silicon\cite{Boykin_TBsp3ds*}. 
The transferability of the parameters is investigated applying them to different 
physical and chemical environments by calculating the bandstructure of two reconstructed 
surfaces with different orientations: Si(100) (2x1) and Si(111) (2x1). The reproduced
$\pi$- and $\pi^*$-surface bands agree in part quantitatively with DFT-GW calculations
and PES/IPES experiments demonstrating their robustness to environmental changes.
We further apply the silicon-parameters to describe the 1D band dispersion of a unrelaxed
rectangular silicon nanowire (SiNW) and demonstrate the EHT-approach of surface passivation
using hydrogen.
Our EHT-parameters thus provide a quantitative model of bulk-silicon and silicon-based 
materials such as contacts and surfaces, which are essential ingredients towards a 
quantitative quantum transport simulation through silicon-based heterostructures.
\end{abstract}

\maketitle

\section{Introduction}
%
Silicon is the dominant component in fabrication of semi-conducting devices
and continues to play a key role in future nanoelectronic devices. 
Recent STM-experiments of molecules  to highly doped silicon (n- or p-type doping) 
showed NDR-behaviour at room temperature in the molecular current-voltage characteristics,\cite{GuisingerNDR} 
which might be a modest initial step towards a molecule based electronics. 
The NDR-behavior, however, was predicted and theoretically demonstrated by Datta and co-workers 
using a physical resonable model for the electronic structure of silicon. Therefore, an correct 
model for the electronic structure of all constituents is essential for a quantitative modeling 
of quantum transport on nano- and atomistic scales.

DFT-based approaches are well-suited to determine electronic and atomic material properties which 
depend on the total energy.\cite{ParrDFT} Abinitio methods in various approximations such as DFT-LDA/GGA 
have been successful to describe properties of molecules and metals,\cite{BarthDFT,MartinElStruct} but they 
are less benchmarked for the electronic properties of semi-conducting materials, particularly for multiple 
bandedges and effective masses - both of them are important components for a quantitative simulation of 
quantum transport. For semi-conductors one well known failure of DFT-LDA/GGA calculations is that the bandgap 
is systematically underestimated by about a factor of $2$,\cite{BernholcNEGFMolSi} which makes a quantitative 
modeling for example of silicon difficult. Quantitative correct bandgaps can be obtained within the 
GW-approximation,\cite{RohlfingGWSiBulk} which is computationally expensive and might thus have limited usage 
at least for transport through large nanostructures. 

At the other end are less rigorous, but computationally less expensive semi-empirical methods
such as orthogonal and non-orthogonal tight binding approaches. Here, the electronic structure
is not rigorously calculated, but determined by an optimization of the free parameters such as the 
matrix elements of the Hamiltonian and overlap to match bandedges and effective masses.
Orthogonal tight-binding (OTB) approaches have been extensively developed in the past and further
optimized to describe the electronic structure of bulk semi-conductors such as silicon and germanium, 
for example\cite{VoglSP3S*,Jancu1_SP3D5S*,Boykin_TBsp3ds*}. 
However, the transferability of the bulk optimized parameters - a problem of any semi-emipirical approach - 
has not been clearly demonstrated, for example, by calculating the electronic structure of (re-constructed) 
surfaces of different orientations.
In turn, their robustness has been mainly shown for atomic structure optimization of bulk, surfaces, and finite size 
clusters, which requires the calculation of the total energy.\cite{ChadiSemSurfMinimEnergy,ChadiSi100Surf,ChadiSi111Surf,
HanssonSi111Surf,ChadiSurfRecon,ChadiSemSurfMinimEnergyII,SwihartClusterSi}
The latter, however, is just an integral property of the band dispersion and eventual errors in the bandstructure 
cancel.\cite{PollmannETB}

In this paper we use a non-orthogonal tight-binding scheme - Extended H\"uckel Theory (EHT) - 
to calculate the electronic structure of the technologically important material silicon.
In our first paper, we applied the EHT approach to carbon nanotubes demonstrating the transferability 
of the EHT-parameters for carbon to small diameter tubes as well as to a strongly deformed CNT-molecule 
heterostructure.\cite{KienleEHTCNT} Here, we present optimized EHT-parameters for bulk-silicon and
benchmark its bandstructure against multiple target values such as bandedges and effective masses. 
We explore the transferability of these EHT-parameters to different environments by calculating 
the 2D-bandstructure for two reconstruced silicon surfaces, silicon (100) (2x1) and (111) (2x1), 
and compare them quantitatively to experiments and state of the art DFT-GW calculations.
Finally, we use the silicon parameters to calculate the 1D band dispersion, density-of-states (DOS), 
and transmission (T) for a un-reconstructed silicon nanowire with and without hydrogen passivation 
demonstrating the capabilities of Extended H\"uckel Theory to model passivated surfaces of nanostructures.
The EHT-parametrization for silicon is shown to be quite transferable to different environments, so that
a quantitative modeling of silicon-based devices becomes feasible.

The paper is organized as follows: section II summarizes briefly the main features of Extended 
H\"uckel Theory. The optimized EHT parameters for bulk silicon along with the comparison of the 
bandedges and effective masses to experimental target values are discussed in section III.
We then investigate the transferability of the parameters by employing them to different 
silicon surfaces. Finally, the 1D electronic structure of a silicon nanowire is determined including 
its surface passivation, and summarize our work in section IV.

\section{Approach}
The silicon bandstructures for the bulk, the two reconstruced surfaces, and the 1D nanowire are calculated 
within a non-orthogonal Slater-Koster scheme\cite{KosterNTB} using Extended H\"uckel Theory to generate the 
overlap- and Hamiltonian-matrix elements ${\bf S}$ and ${\bf H}$, respectively. Here, we briefly summarize 
the essential features of EHT, which is described in more detail in Refs.\cite{Murrell1972,KienleEHTCNT}.

Extended H\"uckel Theory is a semi-empirical theory to calculate the electronic structure of molecules and 
elemental solids. The most striking difference between EHT and orthogonal tight-binding (OTB) is that in EHT 
one works with explicit atomic-like orbital basis functions (AO), which are used to construct the matrix elements 
${\bf S}$ and ${\bf H}$. In turn, in orthogonal-tight binding the basis functions are not known and used
as a formal tool to construct all matrix elements of the Hamiltonian. The matrix elements are then usually
adjusted to a reference bandstructure, for example. Compared to OTB in Extended H\"uckel Theory one adjusts
only the diagonal matrix elements of the Hamiltonian (onsite energies) and the parameters specifying the
basis functions, which are Slater type functions (STO).\cite{CerdaEHT,KienleEHTCNT} 
Since the basis functions are known, the overlap matrix ${\bf S}$ is calculated explicitly and used to 
construct the off-diagonal matrix elements of the Hamiltonian (hopping) according to\cite{Murrell1972,KienleEHTCNT}
\begin{eqnarray} \label{EHT_Rule}
H_{\mu\mu} &=& E_{\mu\mu}~, \nonumber\\
S_{\mu\nu} &=& \int d^3 {\bf r}~\phi_{\mu}^{*} ({\bf r})~\phi_{\nu} ({\bf r})~,\nonumber \\
H_{\mu\nu} &=& \frac{1}{2} K_{eht}~S_{\mu\nu} \left( H_{\mu\mu} 
+ H_{\nu\nu}\right)~,
\end{eqnarray}
assuming that the Hamiltonian depends linearly on the overlap.\cite{Murrell1972}
The original EHT-prescription, cf. Eq.~(\ref{EHT_Rule}), can be further generalized, so that heterogeneous systems 
such as heterostructures and interfaces can be modeled.\cite{KienleMetalCNT,KienleEHTCNT,RazaSTMBond}
The labels $\mu,\nu$ refer to the atomic orbitals, and $S_{\mu\nu}$ is the overlap matrix between the orbital basis 
function $\phi_{\mu}$ and $\phi_{\nu}$, respectively. $K_{eht}$ is an additional fit parameter usually set to $1.75$ 
for molecules and $2.3$ for solids.\cite{Murrell1972,CerdaEHT}
%
Compared to the formal OTB-basis set, Slater basis functions are non-orthogonal, i.e. ${\bf S}\neq {\bf I}$,
which provides an improved transferability of the model parameters with respect to changes in the
environment\cite{MartinElStruct,GoringeRepTB,PecciaTrans}; the enhanced transferability can be justified by constructing 
orthogonal L\"owdin orbitals from the non-orthogonal basis functions. Compared to the original AO's these L\"owdin 
functions are more long-ranged to enforce orthogonality among different L\"owdin functions over the entire domain. 
Consequently, the L\"owdin basis becomes more sensitive to changes in the actual environment.

A concrete example where the transferability of parameters becomes evident are structural deformations.
For bulk-like systems empirical scaling rules have been developed,\cite{KeatingStrain} so that effects of strain on the 
electronic structure can be studied using an orthogonal tight-binding approach employing the commonly used nearest 
neighbor approximation (NNA). 
Even though these empirical approaches for structural deformations are calibrated and work well for bulk systems, 
the transferability of the bulk-optimized parameters along with the scaling rules have not been tested and benchmarked, 
for example, for reconstructed silicon surfaces.
The difficulty in modeling (reconstructed) surfaces within OTB is that surface reconstruction 1). is usually 
accompanied by large structural changes beyond $2-5\%$ and 2). is accompanied by bond changes. 
With respect to point 1). it is not {\it a priori} obvious whether the scaling rules remain valid beyond small 
deformations for which they have been calibrated; 
regarding point 2). it is questionable if a nearest-neighbor approximation which works for bulk can be consistently 
applied to reconstructed surfaces. Using empirical scaling rules the hopping matrix elements between neighboring
atoms of the deformed structure are usually determined from the bulk hopping matrix element between the same atoms.
In the case of the two dimer atoms the problem is that their initial bulk hopping matrix elements is zero, since 
their distance with $\approx 3.8$ \AA is larger than the nearest-neighbor cut-off of $2.35$ \AA.
However, to reproduce the experimentally observed $pi$- and $\pi^*$ surface bands it is important that the electronic
structure model provides the correct hopping matrix elements between the two dimer atoms.

\section{Results for EHT-Electronic Structure for Silicon}
%
\subsection{Si-Bulk}
To perform quantitative transport calculations through nanostructure materials, the free parameters 
of a semi-empirical tight-binding model have to be calibrated to experimental targets and/or 
bandstructure data obtained from other theoretical approaches.
For Extended H\"uckel Theory Cerda and Soria have developed EHT-parameters for several bulk 
crystal structures such as metals, semi-conductors, and compounds.\cite{CerdaEHT} 
Specifically, for silicon these parameters have been optimized to match the bulk-dispersion of
DFT-GW calculations of Rohlfing {\it et al.}\cite{RohlfingGWSiBulk} at selected points within 
the 3D Brillouin zone.\cite{CerdaEHT}

We start by optimizing the EHT-parameters using experimentally determined bandstructure characteristics 
of bulk silicon such as bandedges and effective masses as targets.\cite{MadelungDataBook1991}
We use the TBGreen code\cite{CerdaWEB} to minimize the root-mean square error (RMS) between our EHT-bands 
and the targets via a conjugate gradient method as described in Ref.\cite{CerdaEHT}.
Since most of the targets refer to experiments done at low temperatures ($5-10$ K\cite{MadelungDataBook1991}), 
we perform the minimization of the RMS-error at $T=0.0$ K. The 3D bandstructure of silicon is calculated using
a $sp^3 d^5$ orbital basis for each silicon atom. In order to capture and to optimize the split-off 
gap $\Delta_0$, the Hamiltonian is spin-dependent through the spin-orbit coupling, $LS_p$, used as additional 
fit parameter.

\begin{figure}[htbp]
\centerline{\epsfxsize=8cm
\epsffile{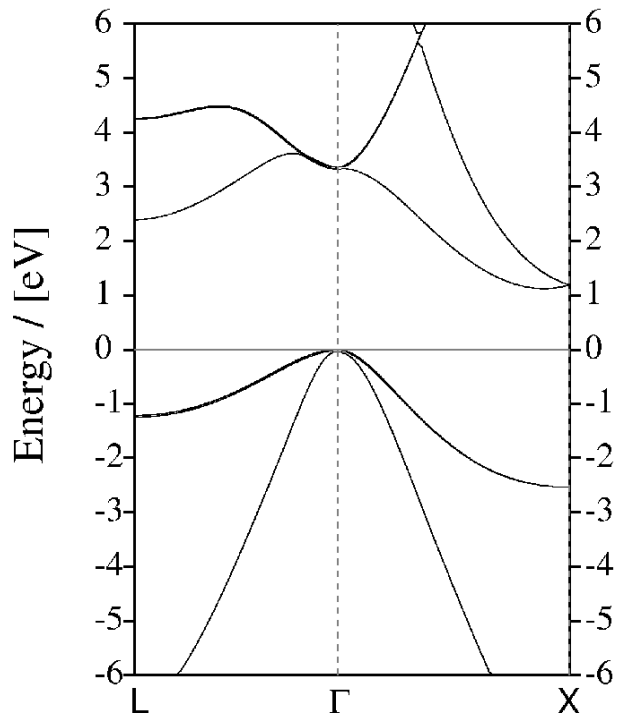}}
\vspace{-0.2cm}
\caption{Bandstructure of bulk-silicon calculated within EHT using 
the parameters given in Table~\cite{Note-Parameters}
The Fermi 
level is at $E_F=0.0$ eV as indicated by the horizontal solid line.
The EHT parameters are optimized to experimental target values taken
from Ref.\cite{MadelungDataBook1991}.} 
\label{Fig_SiliconBulkEK}
\end{figure}

Figure~\ref{Fig_SiliconBulkEK} shows the band dispersion for bulk-silicon using the optimized EHT-parameters 
for a silicon atom.\cite{Note-Parameters}
%

In Table~\ref{TableEffMass} we show the final results of our optimization along with
their relative error (2nd and 3rd column) with the experimental target values as
reference (3rd column). Our values for the bandedges as well as for the effective masses
agree very well with the target values,\cite{MadelungDataBook1991} and are in their quality
competitive to the state-of-the art orthogonal $sp^3sd^5 s^*$ tight-binding model for 
bulk-silicon.\cite{Boykin_TBsp3ds*}
%
%
%
%
%
\begin{table*}
\begin{tabular}[t]{ c | c | c | c | c | c } 
\hline\hline
Quantity & Si-EHT & Rel.Err. $[\%]$ & Si-Target & Si-$sp^3 d^5 s^*$\cite{Boykin_TBsp3ds*} & Rel.Err. 
$\%$\cite{Boykin_TBsp3ds*} \\ \hline
$E_c^{\Gamma}$      &  $3.324$    &    $1.3$  &   $3.368$    &  $3.999$   &  $0.9$   \\
$E_v^{\Gamma}$      &  $0.0$      &    $0.0$  &   $0.0$      &  $0.0$     &  $0.0$   \\
$\Delta_0$          &  $0.0445$   &    $1.0$  &   $0.045$    &  $0.0472$  &  $4.9$   \\
$E_{c,min}^{L}$     &  $2.393$    &    $0.3$  &   $2.400$    &  $2.383$   &  $0.7$   \\
$E_{c,min}^{X}$     &  $1.122$    &   $-0.4$  &   $1.118$    &  $1.131$   &  $1.2$   \\
$k_{min}^{[001]}$   &  $88.0\%$   &   $-3.5$  &   $85.0\%$   &  $81.3\%$  &  $4.4$   \\
$m_{X,l}^{(e)}$     &  $0.939$    &   $-2.5$  &   $0.916$    &  $0.891$   &  $2.7$   \\
$m_{X,t}^{(e)}$     &  $0.161$    &   $15.5$  &   $0.190$    &  $0.201$   &  $5.8$   \\
$m_{L,l}^{(e)}$     &  $1.136$    &   $43.2$  &   $2.000$    &  $3.433$   &  $71.7$  \\
$m_{L,t}^{(e)}$     &  $0.140$    &  $-39.7$  &   $0.100$    &  $0.174$   &  $74.0$  \\
$m_{lh}^{[001]}$    & $-0.182$    &   $10.7$  &   $-0.204$   &  $-0.214$  &  $4.9$   \\
$m_{lh}^{[110]}$    & $-0.148$    &   $-0.7$  &   $-0.147$   &  $-0.152$  &  $3.4$   \\
$m_{lh}^{[111]}$    & $-0.149$    &   $-7.0$  &   $-0.139$   &  $-0.144$  &  $3.6$   \\
$m_{hh}^{[001]}$    & $-0.277$    &   $-0.9$  &   $-0.275$   &  $-0.276$  &  $0.4$   \\
$m_{hh}^{[110]}$    & $-0.579$    &    $0.0$  &   $-0.579$   &  $-0.581$  &  $0.3$   \\
$m_{hh}^{[111]}$    & $-0.663$    &   $10.2$  &   $-0.738$   &  $-0.734$  &  $0.5$   \\
$m_{so}$            & $-0.217$    &    $7.1$  &   $-0.234$   &  $-0.246$  &  $5.1$   \\
$E^G_1$             & $-12.11$    &    $3.1$  &   $-12.50$   &     -      &     -    \\
$m^G_1$             & $ 1.77$     &  $-47.7$  &   $1.20$     &     -      &     -    \\ 
\hline\hline
\end{tabular}
\caption{Bandstructure characteristics for bulk silicon using Extended H\"uckel Theory
and spd-orbitals for each Si-atom. The EHT-parameters, cf. Table\cite{Note-Parameters} 
have been optimized to experimental target values\cite{MadelungDataBook1991} (4th column). 
The last two column on the right are the fit values and errors based on an orthogonal
tight-binding model using $sp^3 d^5 s^*$ orbitals.\cite{Boykin_TBsp3ds*} 
The effective masses at the $L$-valley are not well established, so that they are
to strongly weighted in the optimization.}
\label{TableEffMass}
\end{table*}
%

\subsection{Si-Surfaces for different Orientations}
%
We now investigate the transferability of our EHT-parameters for bulk-silicon and use them
to calculate non-selfconsistently the surface bandstructure for two different surface orientations.
The semi-infinite surfaces are modeled by a finite slab consisting of a series of layers.
Because of this truncation one introduces an additional free surface at the bottom of the slab 
and hence unphysical surface states, which can extend towards the true surface and thus tamper the 
surface band dispersion. To eliminate the dangling bond states at the bottom of the slab we passivate 
in each case the surface of (100) and (111) orientation by attaching hydrogen.

\begin{figure}[htbp]
\centering{
\includegraphics[angle=-90,width=7cm]{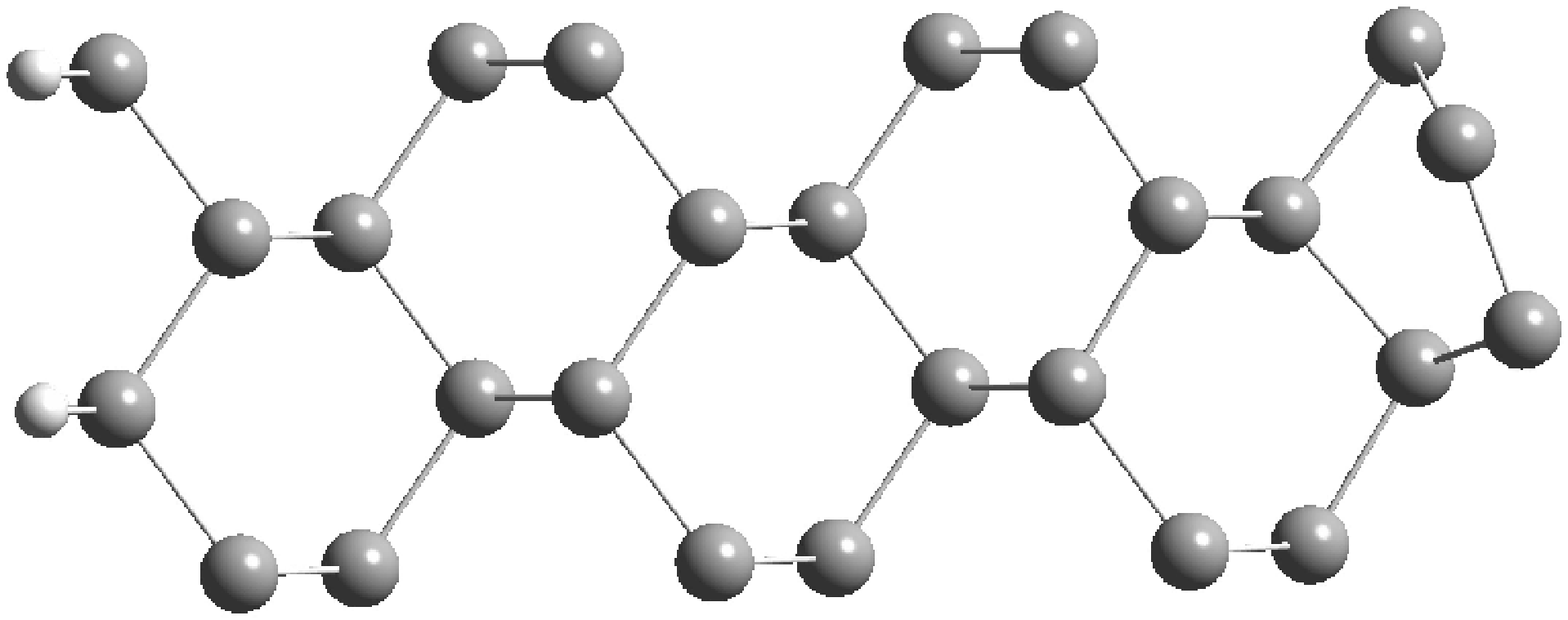}
\vspace{0.5cm}
\includegraphics[angle=-92,width=7cm]{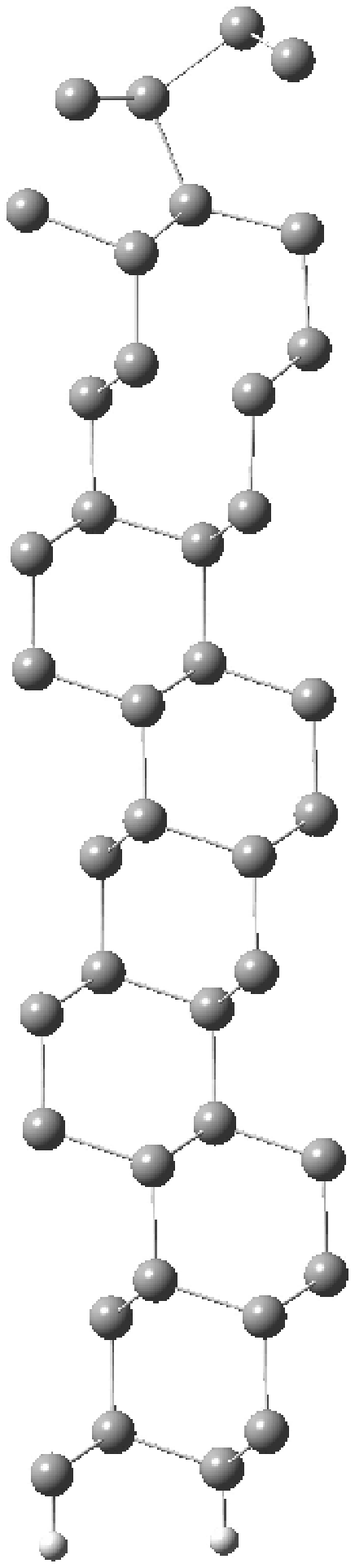}}
\vspace{-0.8cm}
\caption{Structure of the unit cell for the two silicon surfaces.\cite{RamstadCoords-Si100,DengCoords-Si111}.
Top:Si(100) (2x1) with the first 4 layers relaxed and 9 bulk-like layers 
Bottom: Si(111) (2x1) where the first 8 layers are relaxed and 12 bulk
layers. In each case, the bottom of the surface is hydrogen passivated to 
remove dangling bond states.}
\label{Fig_StructureSlabSilicon}
\end{figure}

To calculate the 2D silicon bandstructure of the reconstructed surface we use the unit cell coordinate 
of Ref.\cite{RamstadCoords-Si100} for Si(100) (2x1) and for Si(111) (2x1) from 
Ref.\cite{DengCoords-Si111} as shown in Figure~\ref{Fig_StructureSlabSilicon}.
For Si (100) (2x1) we use a slab with in total 13 layers, where the first 4 layers correspond to 
the reconstructed silicon surface, and the remaining 8 layers correspond to positions of bulk silicon.
The last layer is hydrogen to passivate the bottom of the slab. The lattice vectors for Si(100) (2x1) 
are ${\bf a}_1 = 7.68\AA~{\bf e}_x$ and ${\bf a}_2 = 3.84\AA~{\bf e}_y$. 
Similarly, the Si(111) (2x1) reconstructed surface contains 21 layers with the first 8 layers being 
relaxed, 12 bulk-like layers, and again the last layer hydrogen passivated. 
The 2D Bravais lattice vectors here are ${\bf a}_1 = 6.65\AA~{\bf e}_x$ and ${\bf a}_2 = 3.84\AA~{\bf e}_y$.
%
%

\begin{figure}[htbp]
\centerline{\epsfxsize=7cm 
\epsffile{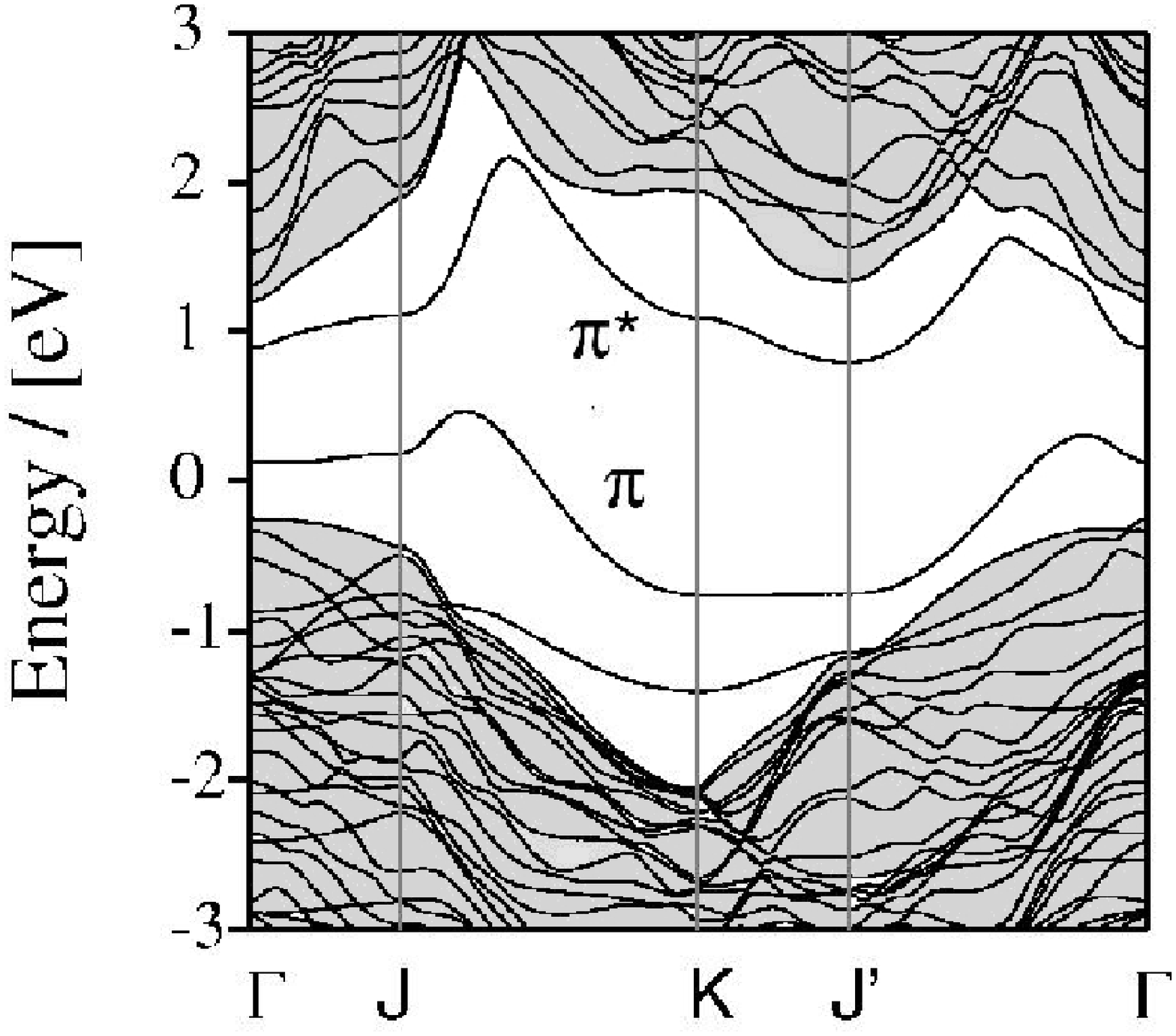}}
\centerline{\epsfxsize=7cm 
\epsffile{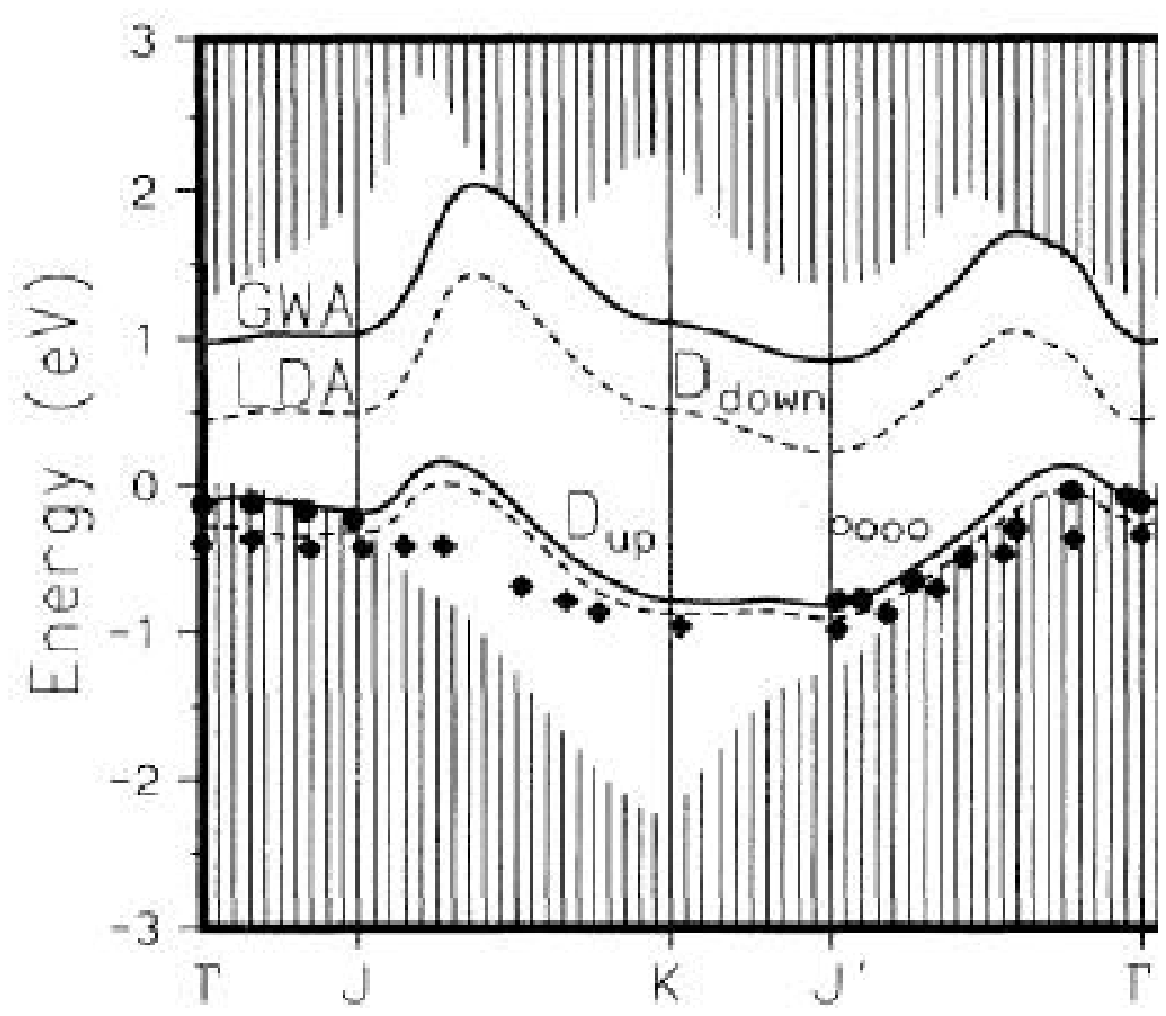}}
\vspace{-0.2cm}
\caption{Surface bandstructure of unpassivated, reconstructed silicon (001) (2x1) calculated within 
EHT (top) using the parameters of Table~\cite{Note-Parameters}
The figure below is a DFT-GW 
calculation\cite{RohlfingSi100GW}. The two bands within the 2D-projected bulk-silicon bandgap correspond 
to the $\pi$ and $\pi^{*}$ bands of the silicon surface described by the ADM.
Reprinted figure (middle) with permission from M. Rohlfing, P. Kr\"uger, and J. Pollmann, {\emph PRB}, 
{\bf 52}, 1905 (1995). Copyright (1995) by the American Physical Society.}
\label{Fig_SiSurf100EK_compare}
\end{figure}

%
Figure~\ref{Fig_SiSurf100EK_compare} shows the bandstructure of reconstructed silicon (100) (2x1) 
calculated within EHT (top) using the silicon parameters of Table~\cite{Note-Parameters}.
The band dispersion at the bottom of Figure~\ref{Fig_SiSurf100EK_compare} correspond to DFT-GW
calculations of Rohlfing {\it et al.}\cite{RohlfingSi100GW} with the $\pi$- and $\pi^*$-bands
in solid lines. As can be seen, our EHT-calculated $\pi$- and $\pi^*$-surface bands agree very well
qualitatively in their shape with DFT-GW calculations as well as with PES experiments for
the $\pi$-band.
%

Despite of the good qualitative agreement, which demonstrates the transferability of our parameters
in capturing the essential physics of the surface bands, there are quantitative differences.
The experimental indirect bandgap $\Delta_{\pi^* - \pi}$ between the $\pi^*$ and $\pi$ band, 
for example, is about $\approx 0.8/0.9$ eV, whereas the gap in EHT is with $\approx 0.3$ eV of 
similar order as in DFT-LDA calculations. 
In Table~\ref{TableEHTSi100} we compare the bandgaps and bandedges calculated in EHT at different 
points of the 2D Brillouin zone with those obtained by DFT-GW calculations\cite{RohlfingSi100GW} 
and experiments.
\begin{table}[htbp]
\begin{tabular}[t]{ c | c | c | c }
\hline\hline
         &  EHT (Table~\cite{Note-Parameters} &  DFT-GW\cite{RohlfingSi100GW}  &  Exp.      \\ 
\hline
$E^{\pi}_{max}$        &  $0.47$   &  $0.16$   &  $-0.42$\cite{Note-DK}                        \\ 
$E^{\pi^*}_{min}$      &  $0.81$   &  $0.81$   &  -                                         \\ 
$\Delta_{\pi^* - \pi}$ &  $0.34$   &  $0.65$   &  -                                         \\ 
$E_{\Gamma}^{\pi}$     &  $0.14$   &  $-0.10$  &  $-0.13$\cite{Note-DK}~,~$-0.42$\cite{Note-DK}   \\ 
$E_{\Gamma}^{\pi^*}$   &  $0.91$   &  $0.94$   &  -                                         \\
$\Delta_{\Gamma}$      &  $0.78$   &  $1.04$   &  -                                         \\
$E^{\pi}_{J}$          &  $-0.20$  &  $-0.19$  &  $-0.26$\cite{Note-DK}~,~$-0.26$\cite{Note-DK}   \\ 
$E^{\pi^*}_{J}$        &  $1.11$   &  $1.00$   &  -                                         \\ 
$\Delta_{J}$           &  $0.91$   &  $1.19$   &  -                                         \\
$E^{\pi}_{K}$          &  $-0.74$  &  $-0.81$  &  $-0.97$\cite{Note-DK}                        \\
$E^{\pi^*}_{K}$        &  $1.11$   &  $1.07$   &  -                                         \\ 
$\Delta_{K}$           &  $1.85$   &  $1.87$   &  -                                         \\
$E^{\pi}_{J'}$         &  $-0.74$  &  $-0.81$  & $-0.81$\cite{Note-DK}~,~$-0.97$\cite{Note-DK}    \\
$E^{\pi^*}_{J'}$       &  $0.81$   &  $0.81$   &  -                                         \\ 
$\Delta_{J'}$          &  $1.55$   &  $1.61$   &  -                                         \\
\hline\hline
\end{tabular}
\caption{Comparison of the bandgaps $\Delta$ and bandedges calculated in EHT, DFT-GW,\cite{RohlfingSi100GW} and
experiments for the silicon surface (100) 2x1 at different points of the 2D Brillouin zone. All values are
in units of eV.}
\label{TableEHTSi100}
\end{table}

\begin{figure}[htbp]
\centerline{\epsfxsize=8cm
\epsffile{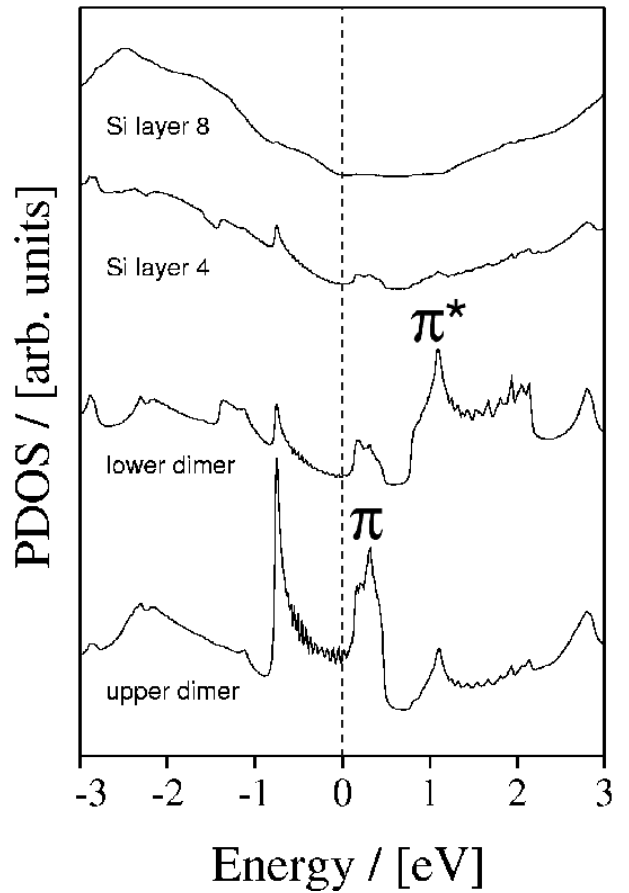}}
\vspace{-0.2cm}
\caption{Density-of-states of un-passivated silicon (100) (2x1) surface calculated for different 
layers starting from the bulk-like 8th layer (top) towards the (100) surface consisting of the
two dimer atoms. The two peaks in the DOS at each dimer atom correspond to the $\pi$- and 
$\pi^*$ bands.}
\label{Fig_SiSurf100DOS_new}
\end{figure}

The differences between the theoretical approaches become more explicit in Figure~\ref{Fig_SiSurf100DOS_new}
where only the E-k dispersion for the $\pi$- and $\pi^*$-surface bands is shown for the EHT- and the DFT-GW 
calculation of Rohlfing {\em et al.}.\cite{RohlfingSi100GW} The plots have been extracted from 
Figure~\ref{Fig_SiSurf100EK_compare} by digitizing the respective $\pi$- and $\pi^*$-bands. 
The unoccupied $\pi^*$-band agrees quantitatively very well with the one obtained from DFT-GW over the 
entire Brillouin zone region as shown in Figure~\ref{Fig_SiSurf100EK_compare}. 
Similarly, the $\pi$-band matches quantitatively DFT-GW as well over a wide range of the Brillouin zone, 
except within the $\Gamma - J$ and the first 3rd of the $J-K$ path. The latter domain includes the
$\pi$-maximum, which appears $\approx 0.25/0.3$ eV too high, so that the $\Delta_{\pi^* - \pi}$ bandgap
is noticably underestimated.
\begin{figure}[htbp]
\centerline{\epsfxsize=8cm 
\epsffile{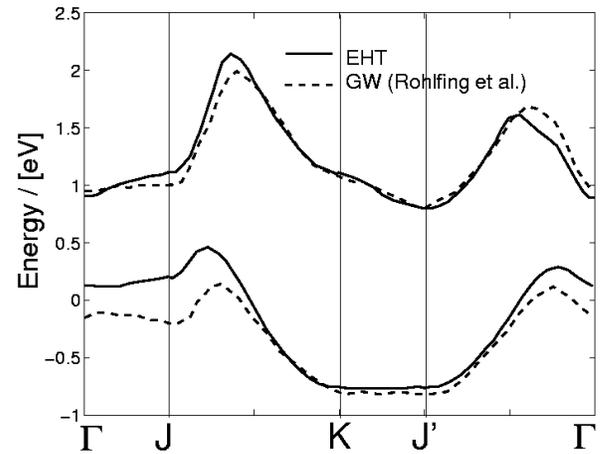}}
\vspace{-0.2cm}
\caption{Comparison of the $\pi$- and $\pi^*$-surface bands calculated within EHT and 
DFT-GW\cite{RohlfingSi100GW}. The dispersion of the $\pi$- and $\pi^*$-bands of 
Figure~\ref{Fig_SiSurf100EK_compare} have been digitized. The data of the red curve are adapted
with permission from Figure 3 of M. Rohlfing, P. Kr\"uger, and J. Pollmann, {\emph PRB}, {\bf 52}, 1905 (1995). 
Copyright (1995) by the American Physical Society.}
\label{Fig_SiSurf100EK_compare2}
\end{figure}

The surface density of states (DOS) is shown in Figure~\ref{Fig_SiSurf100DOS_new}. The energy-resolved partial 
DOS is calculated for each dimer atom (upper and lower) and for the two deeper silicon layers away from the 
surface. The partial DOS of the upper dimer atom is located more closely to the valence band, whereas for the 
lower one it is near the conduction band, indicating that the $\pi$-surface band is formed from the upper-dimer 
atom, whereas the $\pi^*$-band comes from the lower one\cite{Srivastava1999}. Consistent with the $\pi$-band 
dispersion of Figures~\ref{Fig_SiSurf100EK_compare} and \ref{Fig_SiSurf100EK_compare2} the PDOS of the upper 
dimer atom is too far away from the valence band, so that the $\pi^* - \pi$ gap in the PDOS is too small. 

Away from the surface and approaching the bulk-like region the weight in the $\pi$- and $\pi^*$-DOS decreases 
continuously (layer 4), and completely disappears once a deeper bulk-like layer is reached (layer 8), so that 
the original bulk-bandgap of $\approx 1.2$ eV is recovered.
%

\begin{figure}[htbp]
\centerline{
\epsfxsize=7.5cm
\epsffile{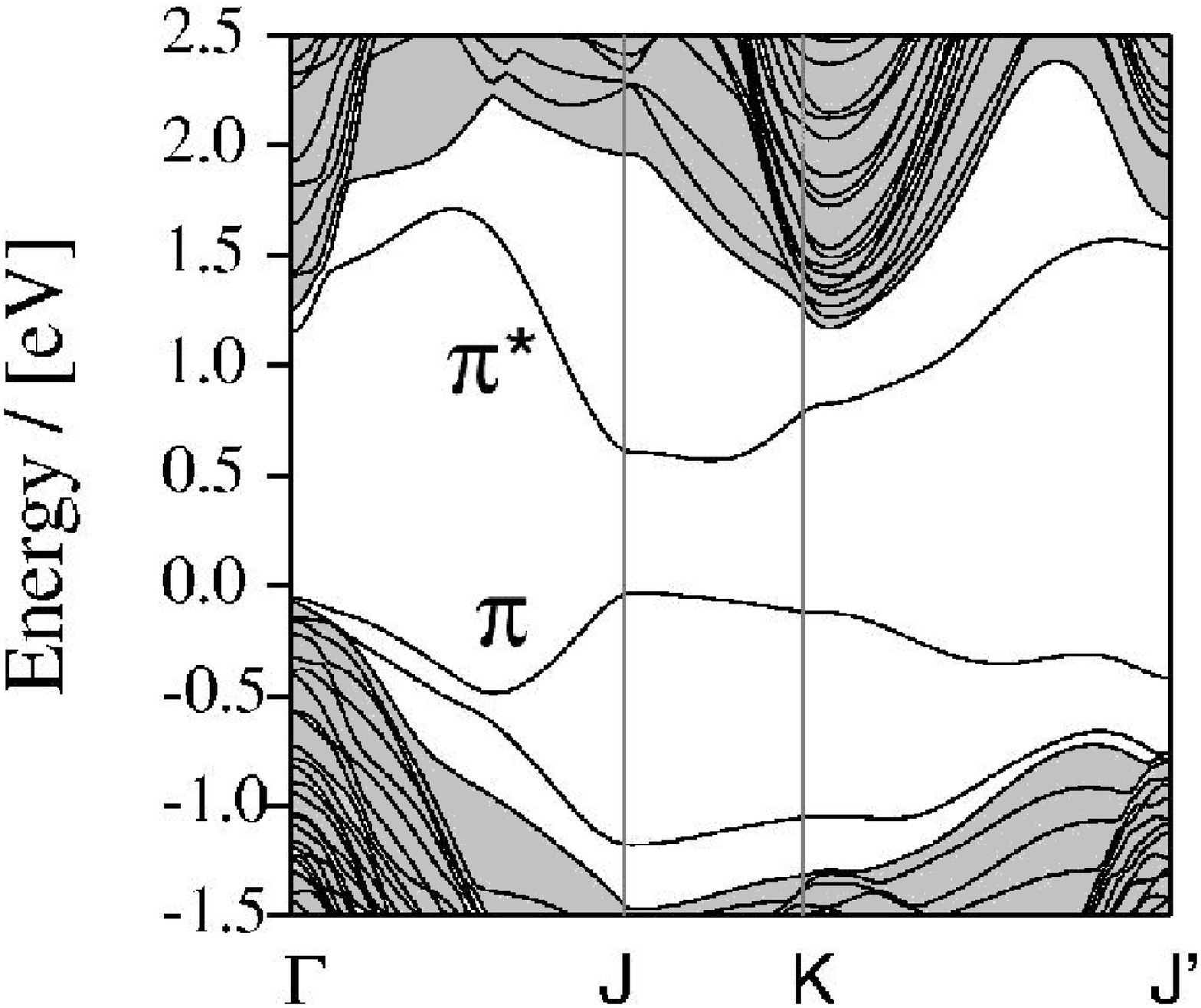}}
\vspace{0.4cm}
\centerline{
\epsfxsize=8.5cm
\epsffile{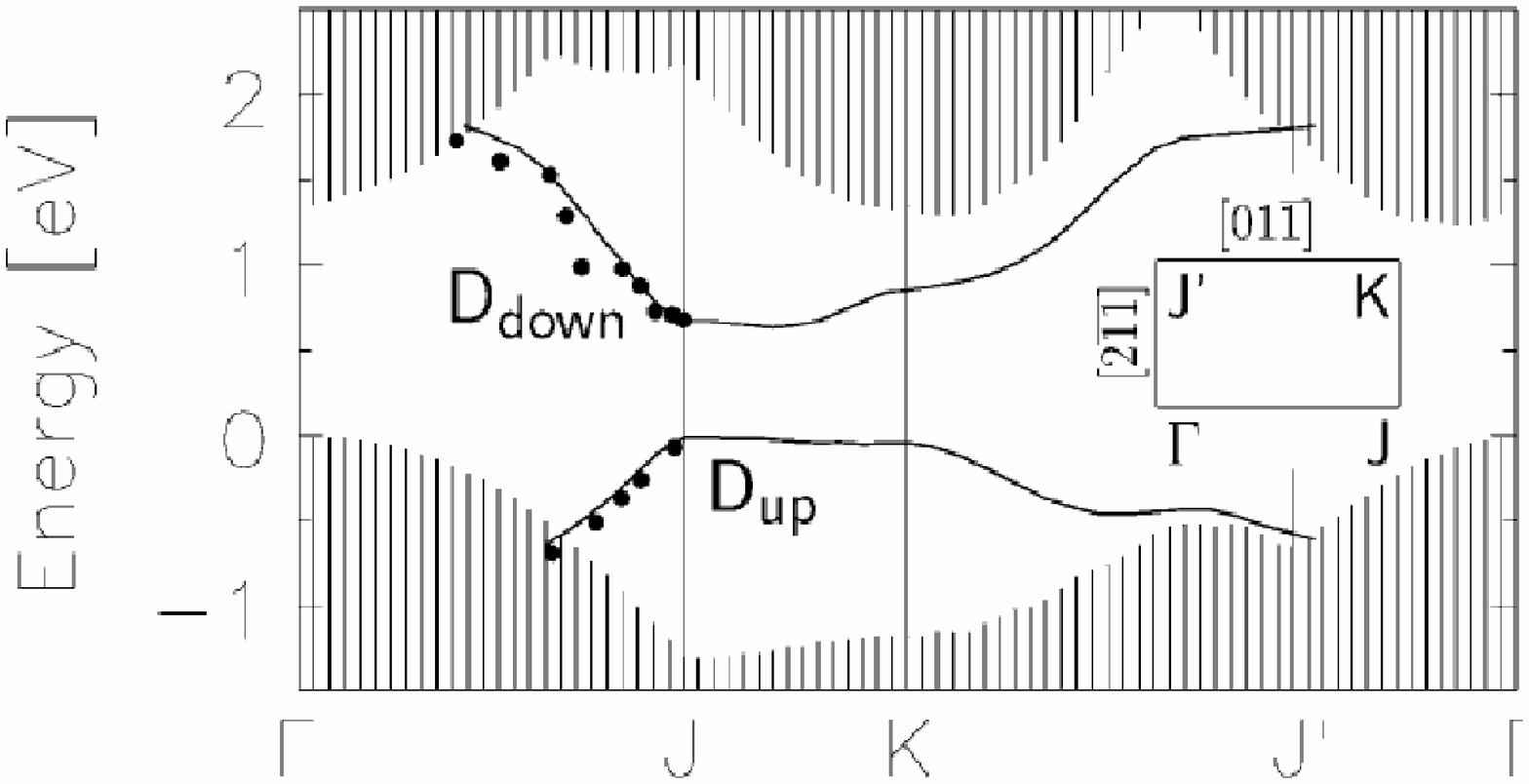}}
\vspace{-0.2cm}
\caption{$\pi$- and $\pi^*$ surface bandstructure of the unpassivated, reconstructed silicon (111) (2x1) 
calculated in EHT (top). The figure at the bottom shows the dispersion calculated within DFT-GGA of Rohlfing 
and Louie\cite{RohlfingSi111_2}, respectively. 
The bottom figure is reprinted with permission from M. Rohlfing and S.G. Louie, {\emph Phys.Stat. Solidi (a)}, 
{\bf 175}, 17 (1999). Copyright (1999) by the American Physical Society.}
\label{Fig_SiliconSurfaceEK111_new}
\end{figure}

\begin{figure}[htbp]
\centerline{
\epsfxsize=8cm
\epsffile{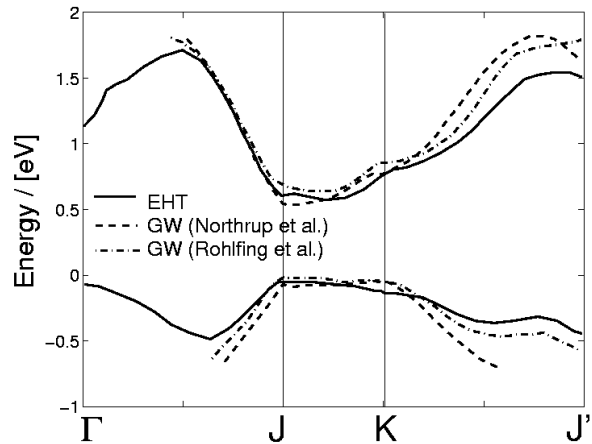}}
\vspace{-0.2cm}
\caption{Comparison of the $\pi$- and $\pi^*$-surface bands calculated within EHT and 
DFT-GW\cite{NorthrupSi111GW,RohlfingSi111_2}. The dispersion of the $\pi$- and $\pi^*$-bands of 
Figure~\ref{Fig_SiSurf100EK_compare} have been digitized. 
The data in the green curve are adapted with permission from Figure 2 of J.E. Northrup, M.S. Hybertsen, 
and S.G. Louie, {\emph Phys.Rev.Lett.}, {\bf 66}, 500 (1991). 
Copyright (1999) by the American Physical Society.
The data in the red curve are adapted with permission from Figure 2 of M. Rohlfing and S.G. Louie, 
{\emph Phys.Stat. Solidi (a)}, {\bf 175}, 17 (1999). Copyright (1999) by the American Physical Society.}
\label{Fig_SiliconSurfaceEK111_compare}
\end{figure}

As second example, we look at the surface bandstructure of reconstructed Si(111) 2x1 as shown 
in Figure~\ref{Fig_SiliconSurfaceEK111_new} using the EHT-parameters in Table~\cite{Note-Parameters}.
Similar to the previous case, the overall shape of the $\pi$- and $\pi^*$-surface bands match
qualitatively well with DFT-GW calculations of Rohlfing {\it et al.}\cite{RohlfingSi111_2}. 
In Table~\ref{TableEHTSi111} we compare our EHT-bandedges and gaps (1st column) at two specific points 
$J$ and $K$ of the 2D-Brillouin zone with DFT-GW calculations\cite{RohlfingSi111_2,NorthrupSi111GW} and 
PES/IPES experiments. The values for the bandedges as well as for the gaps agree quantitatively well 
among all three calculations, and show also a good agreement with PES/IPES experiments, where the error 
in the energy resolution is typically $150-200$ meV depending on temperature and incident energy
of the electrons.
A more extended comparison with PES/IPES-experiments turns out to be very limited, since the $\pi$- and 
the $\pi^*$-bands are not as well experimentally determined as in the case of silicon (100) (2x1). 
Contrary to the previous case of Si(100) (2x1), we find for Si(111) (2x1) that both $\pi^*$- and $\pi$-band
calculated in EHT agree quantitatively very well with DFT-GW calculations of 
Northrup {\it et al.}\cite{NorthrupSi111GW} and in particular with Rohlfing {\it et al.}\cite{RohlfingSi111_2} 
over the entire range of the Brillouin zone as shown in Figure~\ref{Fig_SiliconSurfaceEK111_compare}. 
\begin{table}[htbp]
\begin{tabular}[t]{ c | c | c | c | c }
\hline\hline
         &  EHT (Table~\cite{Note-Parameters}) &  DFT-GW\cite{NorthrupSi111GW}  &  DFT-GW\cite{RohlfingSi111_2} 
& Exp. \\ \hline
$\Delta_{J}$           &  $0.66$   &  $0.60$   &  $0.68$   &    -     \\
$\Delta_{K}$           &  $0.91$   &  $0.82$   &  $0.92$   &    -     \\  
$E^{\pi}_{J}$          &  $-0.05$  &  $-0.07$  &  $0.0$    &  $0.09$  \\ 
$E^{\pi^*}_{J}$        &  $0.61$   &  $0.53$   &  $0.69$   &  $0.67$  \\ 
$E_{K}^{\pi}$          &  $-0.11$  &  $-0.04$  &  $-0.04$  &    -     \\ 
$E_{K}^{\pi^*}$        &  $0.80$   &  $0.78$   &  $0.88$   &    -     \\ 
\hline\hline
\end{tabular}
\caption{Comparison of the bandgaps $\Delta$ and bandedges (in units of eV) calculated in EHT, 
DFT-GW\cite{NorthrupSi111GW,RohlfingSi111_2}, and experiments for the silicon surface 
(111) (2x1) at different points of the 2D Brillouin zone.}
\label{TableEHTSi111}
\end{table}
%

%
In the two previous cases we explored the transferability of the EHT-parameters, 
cf. Table~\cite{Note-Parameters}
, optimized for bulk-silicon by applying them to other environments 
such as re-constructed surfaces for silicon (100) (2x1) and (111) (2x1). 
Without any re-parametrization the experimentally observed $\pi$- and $\pi^*$-surface bands are 
reproduced qualitatively in their overall shape, and in the case of silicon (111) (2x1) we also 
achieve a good quantitative match compared to PES/IPES experiments and DFT-GW calculations as well.
As discussed, quantitative differences exist, particularly for the indirect bandgap $\Delta_{\pi^* - \pi}$ 
for silicon (100) (2x1), which is underestimated similar to DFT-LDA pseudopotential 
calculations.\cite{Srivastava1999}

One reason for the obvious quantitative differences, particularly the wrong position of 
the $\pi$-band above the valence might be due to the non-self consistent calculation of
the bandstructures for the reconstructed silicon (100) and (111) surfaces. 
Calculating the total non-self consistent charge on each dimer atom by integrating the LDOS 
gives a charge of $4.13$e on the upper and $3.75$e on the lower dimer atom. The total charge
of the two dimer atoms is about $7.88$e. In turn, a self-consistent calculation of the dimer 
atom charge using SIESTA,\cite{OrdejonSiesta,SolerSiesta} results in a total charge on the upper 
dimer atom of about $4.0$e, whereas the lower one has $3.88$e. Note, that in the EHT non-scf as 
well as in the SIESTA scf-case the total charge on the two-dimer system is the same and is effectively 
positive with respect to their total number of valence electrons of $8$. 
Qualitatively, what one would expect is that the upper dimer atom which carries initially too much
negative charge, looses parts of it under self-consistency and charge is transfered partly to
the lower dimer atom making, which becomes in turn more negative after self-consistency.
The overall effect of self-cinsistency is then mainly to redistribute charge among the two dimer 
atoms; the initial $\pi$-band which consist of the upper (less negative) dimer would float down, 
whereas the $\pi^*$-band would float up since the lower dimer atom is more negative. Both bands
float in opposite directions, so that the indirect bandgap $\Delta_{\pi^* - \pi}$ increases.

%
\subsection{Si-Nanowire $\langle 100 \rangle$: H-Passivation in EHT}
\begin{figure}[htbp]
\vspace{0.3cm}
\centerline{
\epsfxsize=4cm
\epsffile{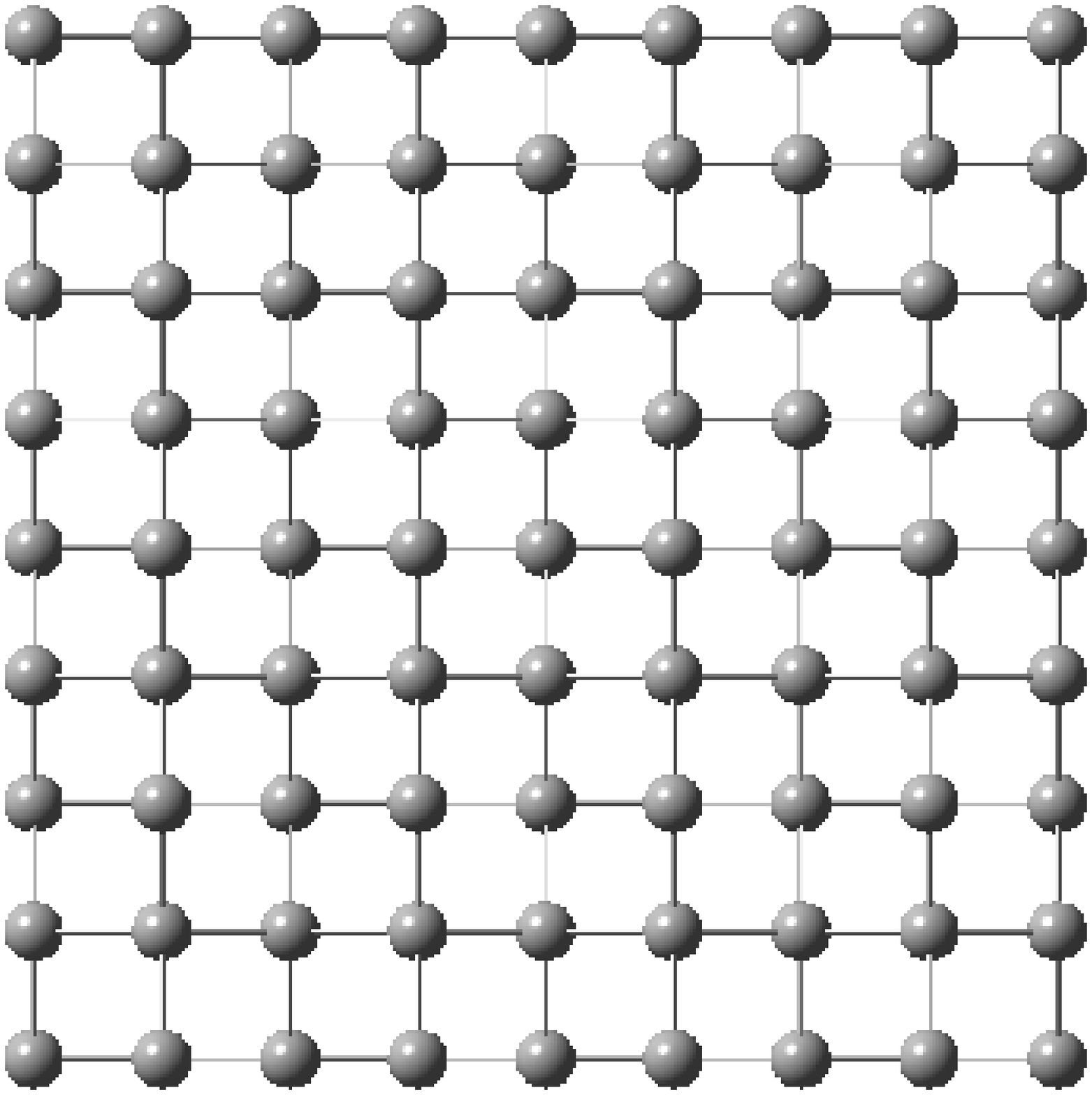}
\hspace{0.3cm}
\epsfxsize=4cm
\epsffile{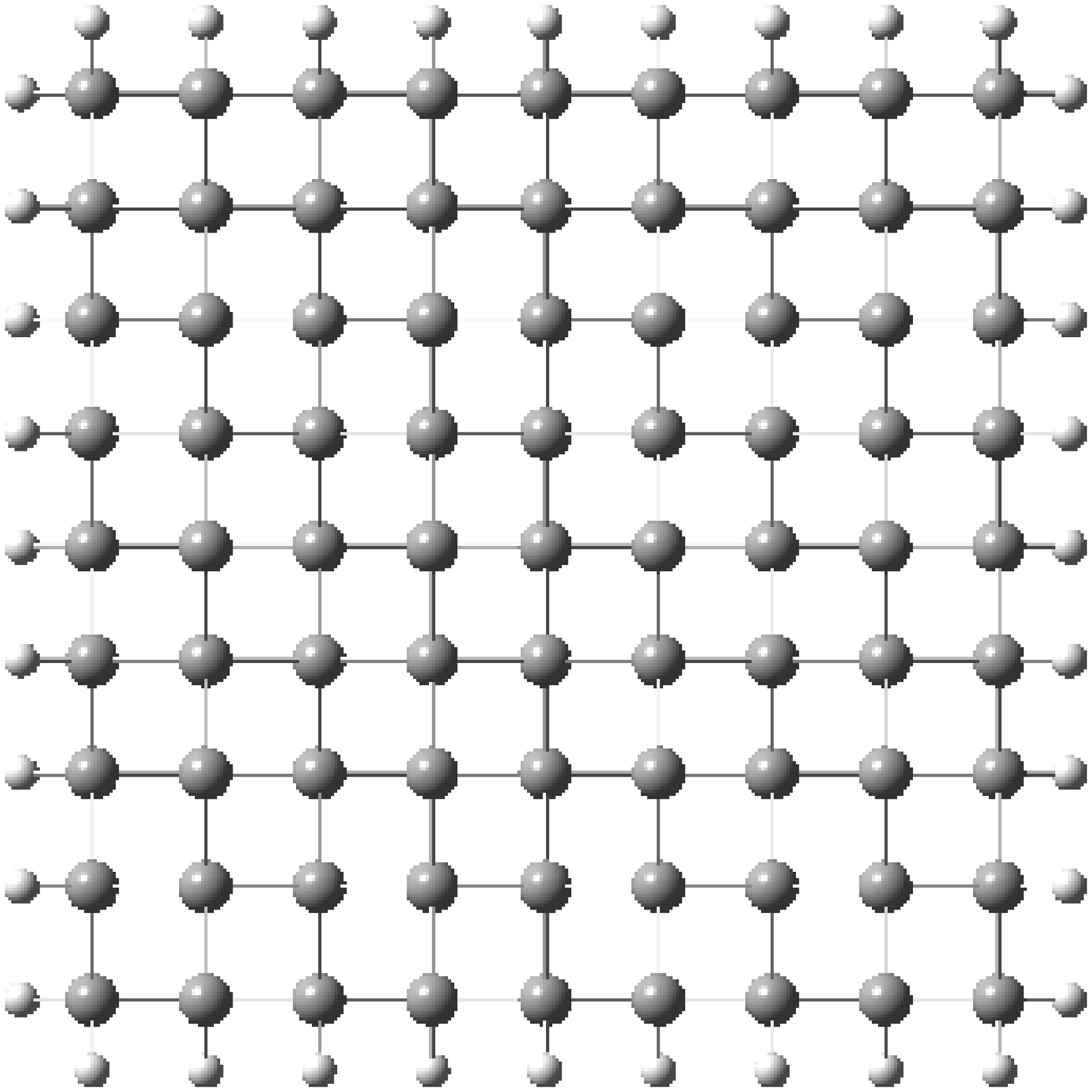}}
\hspace{0.5cm}
\centerline{
\epsfxsize=4cm
\epsffile{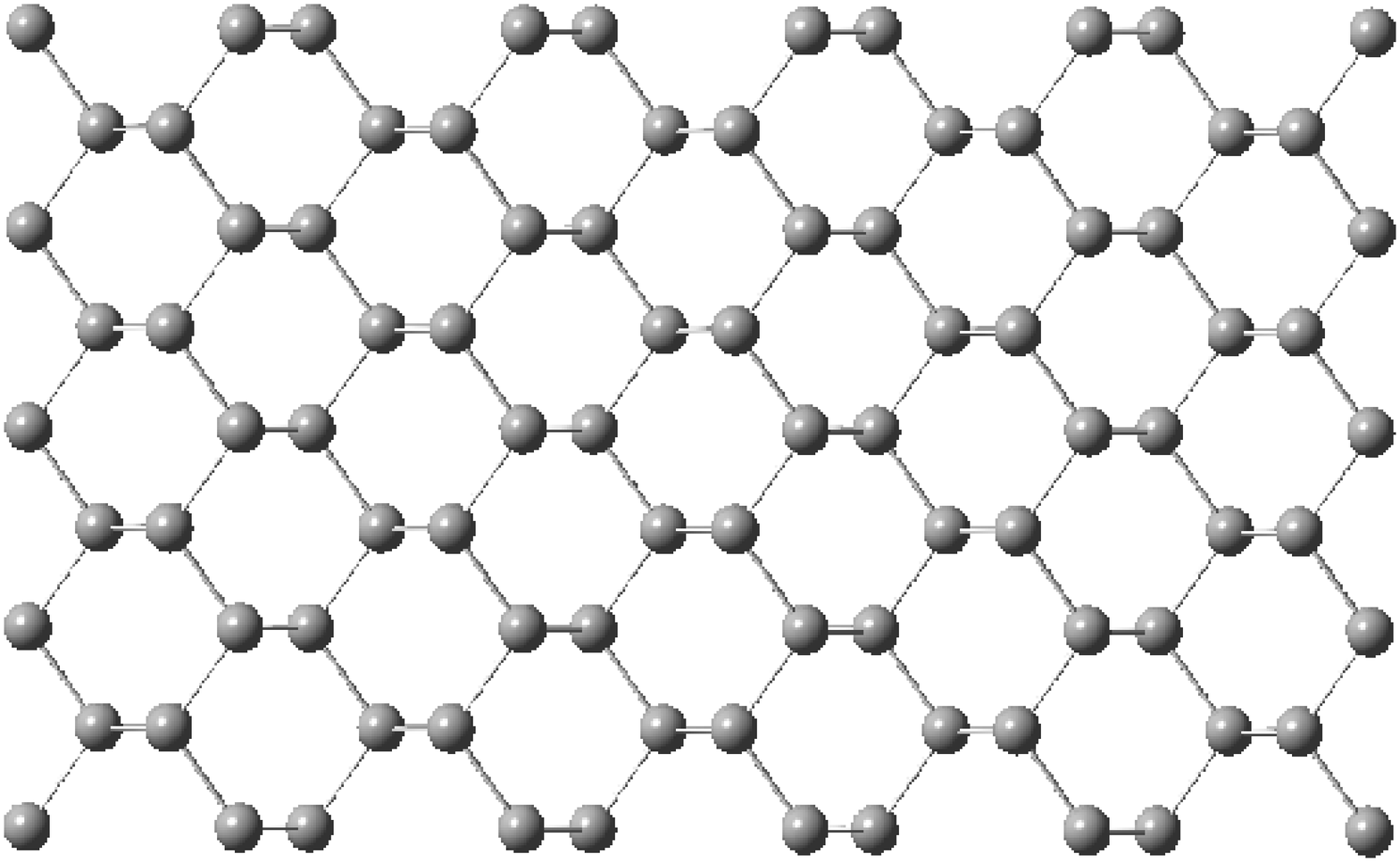}
\hspace{0.3cm}
\epsfxsize=4cm
\epsffile{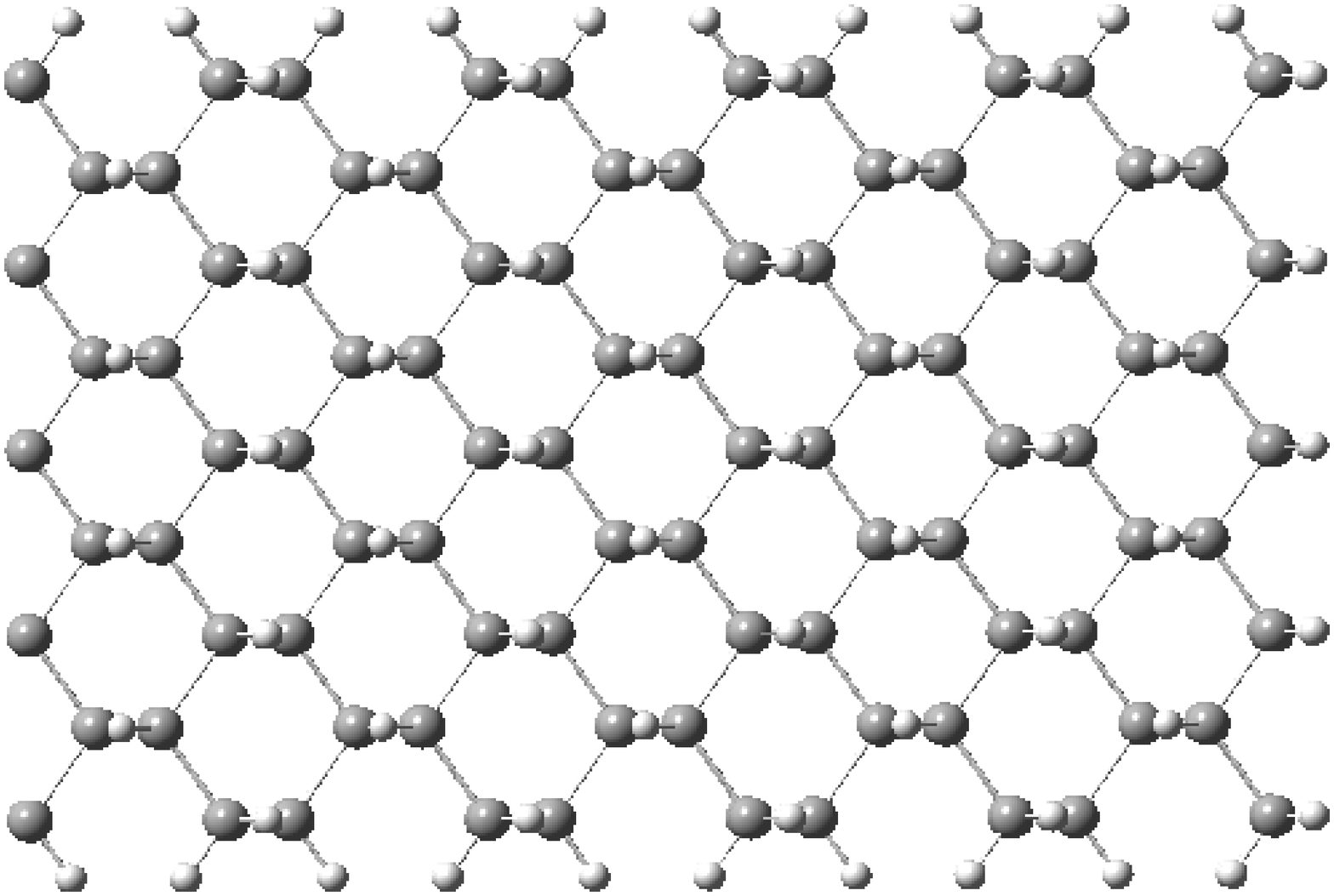}}
\caption{Front (top) and sideview (bottom) of a silicon nanowire with rectangular 
cross section and sidelength $D=1.5$ nm along the $\langle 100 \rangle$ direction. 
For the un-passivated wire (left) the unit cell contains 81 silicon atoms, and in 
the H-passivated case (right) the total number of atoms is 117.}
\label{Fig_Sketch_SiNW}
\end{figure}

As final example, we use the silicon parameters, cf. Table~\cite{Note-Parameters}
in combination with the ones for hydrogen to demonstrate how to passivate 
surfaces of nanostructures in EHT by means of a silicon nanowire. We use a wire
along the $\langle 100 \rangle$ direction with rectangular cross section and
sidelength $D=1.5$ nm as shown in Figure~\ref{Fig_Sketch_SiNW}.
For simplicity we assume that the wire is un-relaxed, i.e. the positions of the
silicon atoms of the wire correspond to the positions of bulk silicon. 
For the surface we consider two cases as illustrated in Figure~\ref{Fig_Sketch_SiNW}: 
i) a wire without H-passivation and ii) a wire where we have explicitly added hydrogen 
atoms to saturate the dangling bonds.

\begin{figure}[htbp]
\centerline{
\epsfxsize=4cm
\epsffile{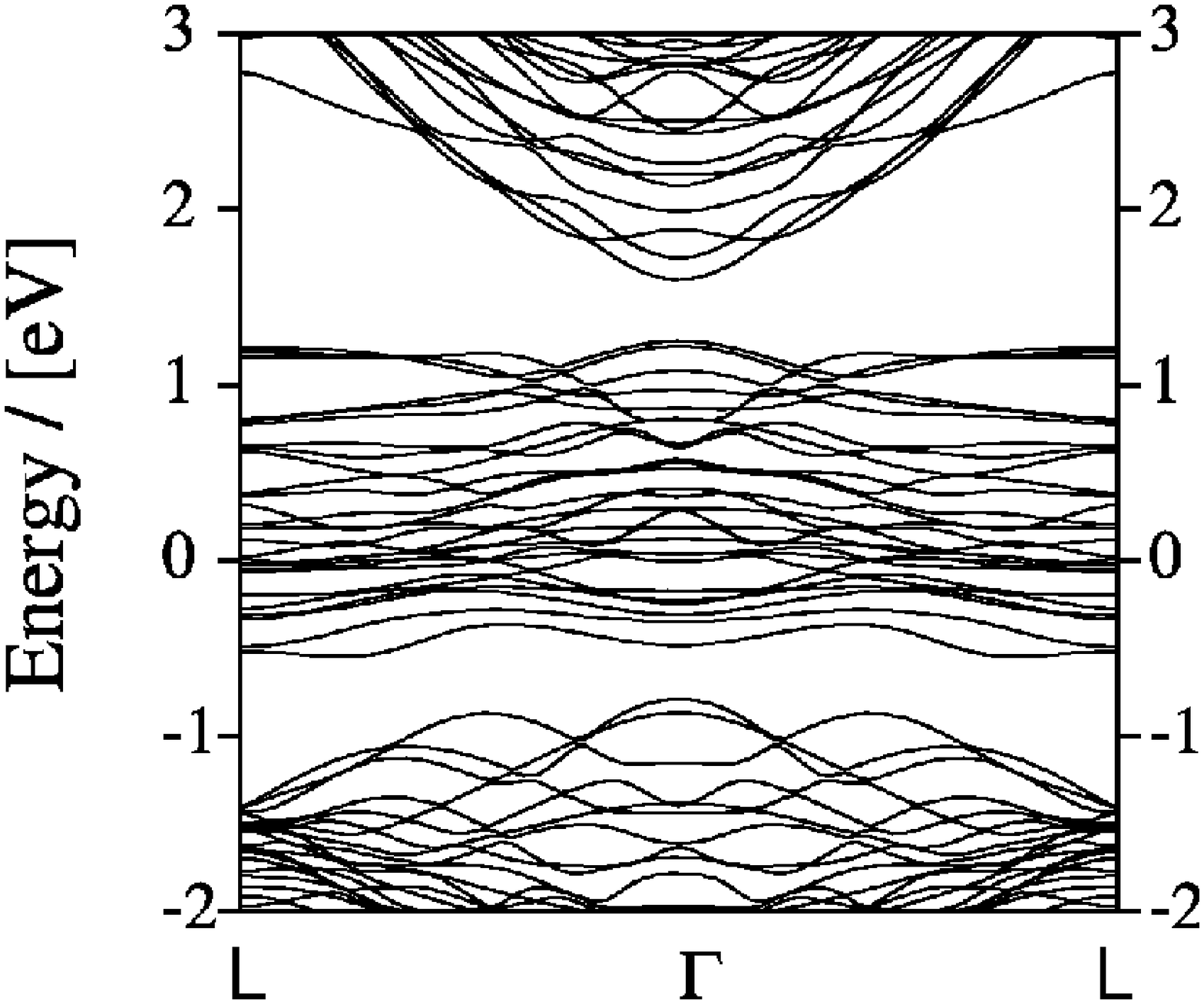}
\epsfxsize=4.3cm
\epsffile{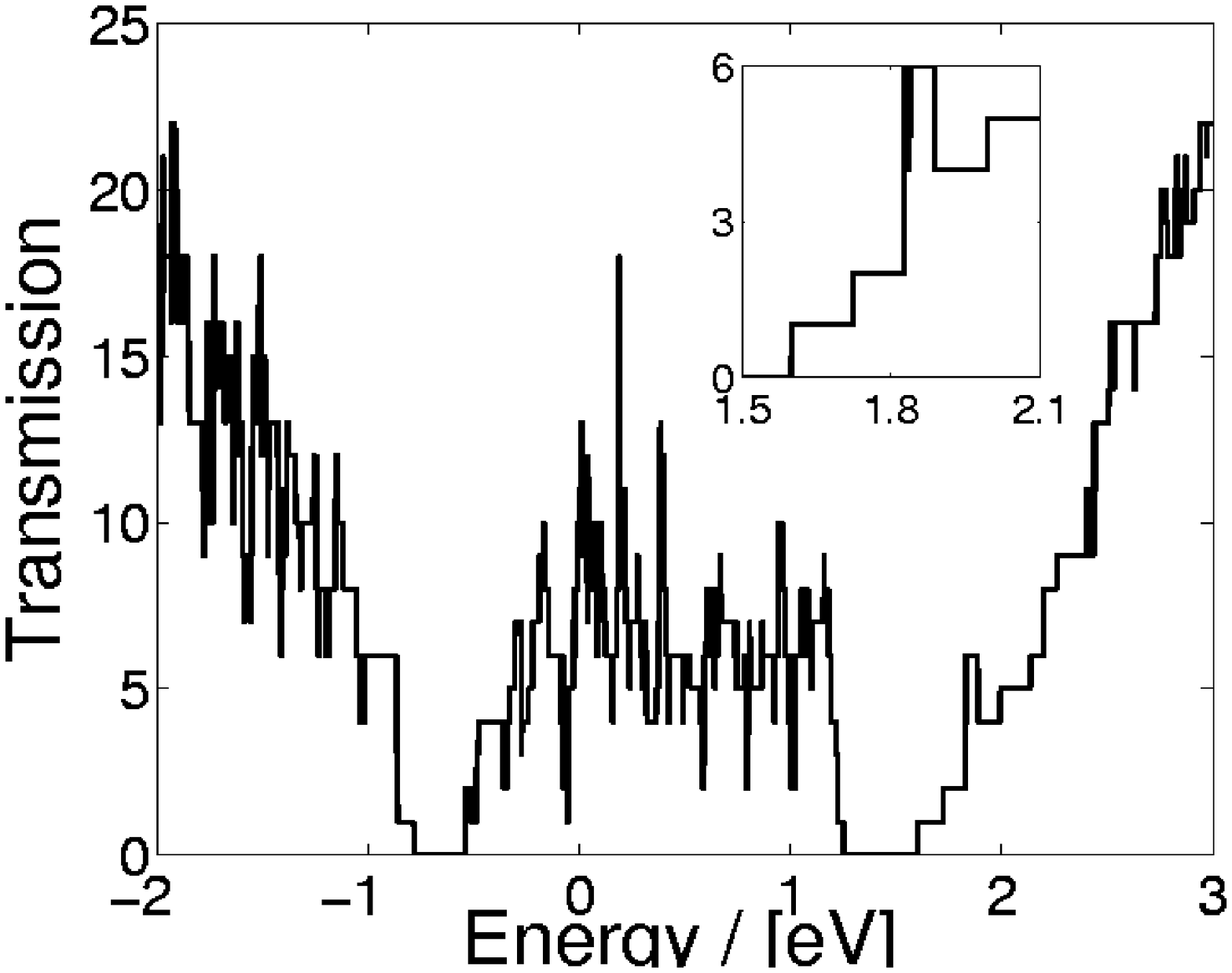}}
\vspace{0.3cm}
\centerline{
\epsfxsize=4cm
\epsffile{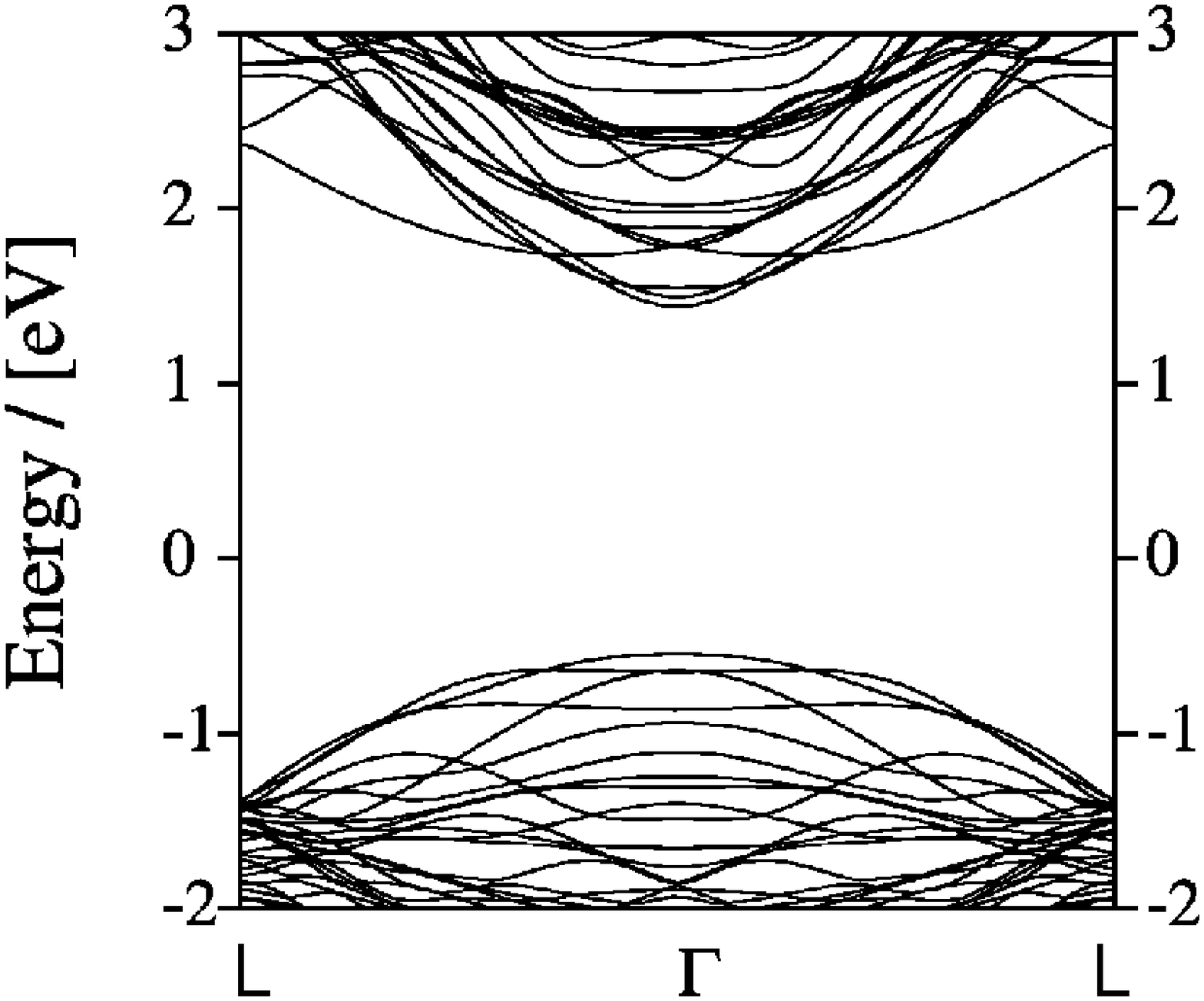}
\epsfxsize=4.3cm
\epsffile{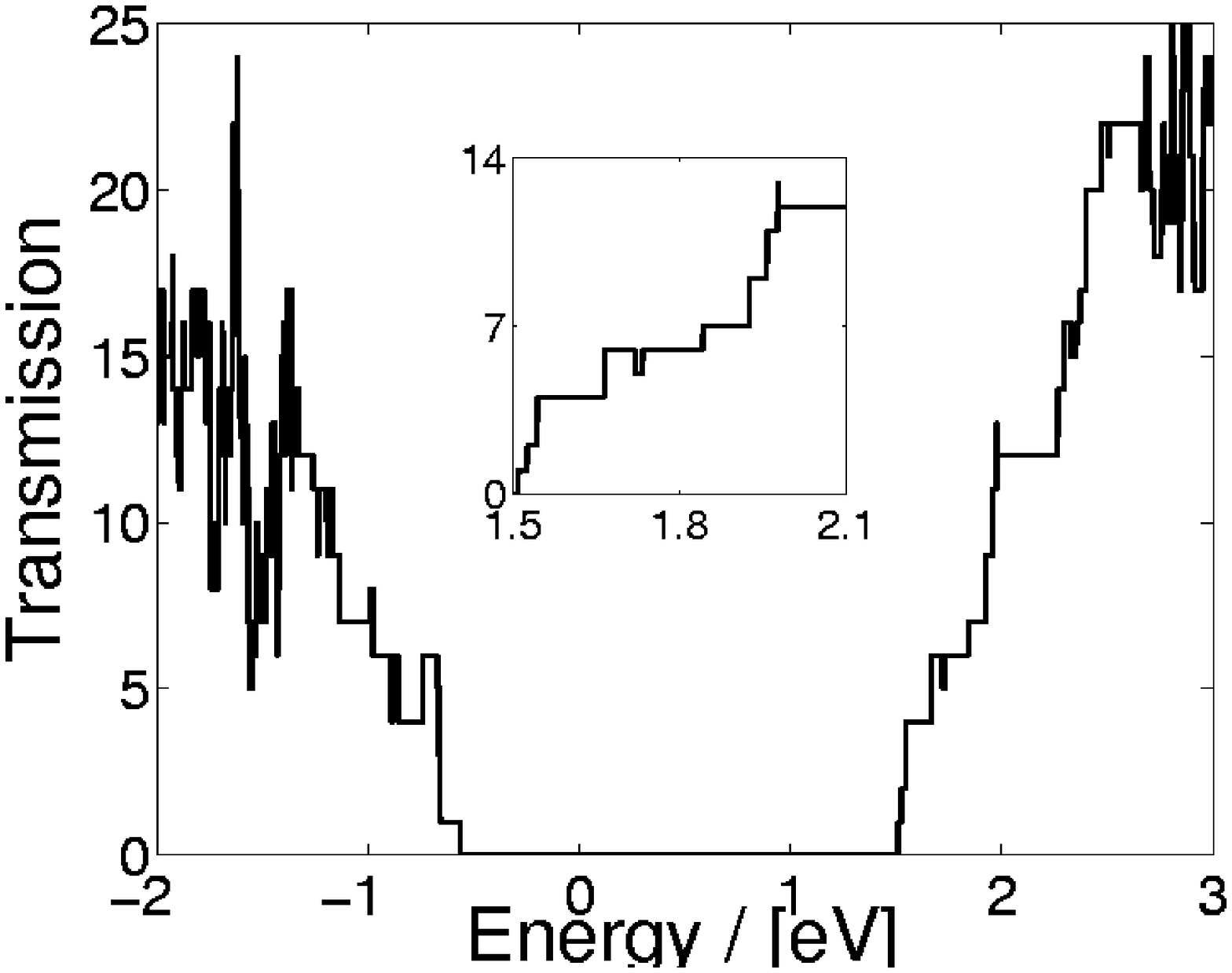}}
\vspace{-0.2cm}
\caption{1D bandstructure and transmission per spin of an un-relaxed silicon nanowire along 
the $\langle 100 \rangle$ wire axis. The $E-k$ and transmission on the left are for the unpassivated 
surface, and on the right the surface is hydrogen passivated to remove the dangling bond states.
The transmission shown in the insets takes integer values where each channel contributes one unit 
quantum conductance $G_0=e^2/h$ per spin.}
\label{Fig_EK1D_SiNW100_Hpass}
\end{figure}

In Figure~\ref{Fig_EK1D_SiNW100_Hpass} the 1D-dispersion relation is shown for the
two structures, cf. Fig.~\ref{Fig_Sketch_SiNW}. As can be seen in the left part, 
the ``bulk''-like bandgap of the silicon nanowire is covered by dangling bond states 
due to the unsaturated bonds of the surface atoms. The entire bands of the 
dangling bonds are completely removed after the wire is passivated by explicitly attaching 
hydrogen (bottom), so that the bulk-bandgap of $\approx 2.1$ eV is recovered. 
Note, that for the case of a H-passivated wire we also assume that the atoms remain in their 
bulk-positions. 

The approach in Extended H\"uckeL Theory to passivate the SiNW surface (and possible other surfaces)
by physically attaching hydrogen atoms, for example, differs from the methods described 
in Ref.\cite{LeeOTBDangBond} within an orthogonal tight-binding scheme. In the latter case, 
the dangling bonds are removed by transforming first the Hamiltonian from a 
$|s \rangle,~|p_x \rangle,~|p_y \rangle,~|p_z \rangle$ representation to a representation using $sp^3$-hybrid 
basis functions. One then has to check which bonds of the surface atoms are not saturated and to increase then 
the respective orbital energy of the dangling bond. In raising the orbital energy manually one formally mimics 
the effect of the hydrogen atoms onto the silicon surface atoms as if hydrogen were attached. The amount, however, 
by which the dangling bond state energy has to be shifted is {\it a priori} not known and has to be determined 
empirically\cite{LeeOTBDangBond}. 

In turn, in Extended H\"uckel Theory the surface passivation is controlled by the passivation atoms, 
which are an explicit part of the entire structure. Once the structure is specified, the amount by which 
the silicon levels are shifted is determined by the chemical species of the passivation atoms.
The hybridization causing the shift of the silicon dangling bond states is naturally incorporated within 
the Hamiltonian matrix through the EHT-prescription, cf. Eq.(\ref{EHT_Rule}).

\section{Summary}
We applied Extended H\"uckel Theory (EHT) to describe the electronic structure 
of several silicon-based structures. A quantitative benchmark of the bulk-silicon
electronic structure was achieved by optimizing the EHT-parameters to experimentally
determined target values specific for the bulk-bandstructure of silicon. 
Our developed parameter set for silicon provides a quantitative good agreement with 
these targets and are competitive with the results based on orthogonal-tight binding for
silicon.

The transferability of the parameter set was investigated by applying them to different
environments such as reconstructed silicon for two different surface orientations. 
Within our EHT-approach using the bulk-optmized parameters we could reproduce the 
essential features of the surface band dispersion, in particular the $\pi$- and $\pi^*$-surface
bands experimentally well established. A quantitative comparison of our EHT-bandstructure with 
DFT-GW calculations as well as PES/IPES experiments shows qualitative and quantitative agreement,
particularly for silicon (111) (2x1) surface. 
However, our indirect bandgap $\Delta_{\pi^* - \pi}$ is lower compared to PES/IPES experiments 
and of the same order as DFT-LDA calculated gaps. This discrepancy is partly due to the non-self 
consistent calculation of the surface electronic structure of reconstructed silicon, where the 
charge redistribution due to structural changes is discarded.
We expect that a full 3D self-consistent solution of the electronic structure, for example 
within a {\it C}omplete {\it N}eglect of {\it D}ifferential {\it O}verlap scheme (CNDO), can
correct for the discrepancies, which would further increase the transferability of our EHT-parameters
for silicon.

Using the silicon nanowire as example we used the EHT-parameters for silicon and demonstrated 
a generic approach to surface passivation of nanostructure materials by physically attaching hydrogen 
atoms to the SiNW surface. The respective dangling bond states are removed from the bandgap region of 
the wire in a systematic manner without the need to shift dangling bond molecular levels by hand.

The flexibility of EHT demonstrated here and in our previous paper (Part I)\cite{KienleEHTCNT}
opens the door to study electronic structure and transport through molecular heterostructures as 
well as larger nanostructures preserving the atomistic features of the system. 
We believe, that Extended H\"uckel Theory is a good practical comprimise between rigorous, 
but computationally expensive DFT-based approaches, and orthogonal tight-binding methods, which 
might not be suitable for large structural deformations beyond $2-3\%$. More coarse-grained models 
such as effective mass might be even prohibitive for the same purpose, since they inherently fail
to account for bonding.
The main appeal of EHT is that it does capture bulk- as well as surface physics along with bonding
chemistry at heterointerfaces and molecular heterostructures - including large structural
deformations - all within a unified semi-empirical framework\cite{KienleEHTCNT,KienleMetalCNT,LiangC60}.
Silicon nanowires, for example, are interesting candidates for new channel materials 
for novel MOSFETs. A widely discussed and still open question is whether structural relaxation of the wire 
significantly affects its electronic structure as well as the overall device performance. 
Intuitively, one expects that the electronic properties particularly of small diameter wires are dictated 
by their surface. Some aspects of this problem is currently investigated where we employ Extended H\"uckel 
Theory to describe the electronic structure of the relaxed wire for different diameters and 
orientations.\cite{LiangSiNW-IV}
Extended H\"uckel Theory as it stands also needs further improvement to fully utilize its capabilities 
and to establish it as a methodological tool towards a quantitative modeling of quantum transport through 
nanostructures. This is left for future work.

\section*{Acknowledgement}
We acknowledge the support of the Army Research Office through the Defense University Research Initiative in 
Nanotechnology (DURINT) program, the Defense Advanced Research Projects Agency-Air Force Office of Scientific 
Research (DARPA-AFOSR). J. Cerda acknowledges support from the Spanish DGICyT under contract 
No. MAT2004-05348-C04-2, respectively. We further acknowledge the Network for Computational Nanotechnology
(NCN) to use the computational facilities. We are indebted to Prof. Tim Boykin for helpful discussions. 
The authors would also like to thank S. Srivastava for helpful discussions.


\begin{thebibliography}{99}
\bibitem{Note-DK}
The experimental values have been extracted from the respective figures.
\end{thebibliography}
\end{document}